\documentclass[prl,twocolumn,aps,floats,epsf,epsfig,amsmath,amssymb]{revtex4}

\usepackage{graphicx}
\usepackage{subfigure} 
\usepackage{dcolumn}
\usepackage{bm}

\begin{document}

\draft

\title{Many-body exchange-correlation effects in graphene }

\author{E.~H.~Hwang,$^1$ Ben Yu-Kuang Hu,$^{2,1}$ and S.~Das Sarma$^1$}
\address{$^1$Condensed Matter Theory Center,
Department of Physics, University of Maryland, College Park,
Maryland  20742-4111}
\address{$^2$Department of Physics,
University of Akron, Akron, OH 44325-4001}

\begin{abstract}
We calculate, within the leading-order dynamical-screening
approximation, the electron self-energy and  spectral function at
zero temperature for extrinsic (or gated/doped) graphene. We also
calculate hot carrier inelastic scattering due to
electron-electron interactions in graphene. We obtain the
inelastic quasiparticle lifetimes and associated mean free paths
from the calculated self-energy. The linear dispersion and chiral
property of graphene gives energy dependent lifetimes that are
qualitatively different from those of parabolic-band
semiconductors.


\end{abstract}




\maketitle

\section{Introduction}

Recent developments in techniques for fabricating conducting
graphene layers \cite{geim1} have provided the physics community
with a unique opportunity to study an interacting two-dimensional
(2D) massless Dirac fermion system. This has led to considerable
experimental and theoretical activity in this field
\cite{geim,kim}. The band structure of graphene, by dint of its
honeycomb lattice, has linear dispersions near the K and K$'$
points of the Brillouin zone \cite{wallace}. The corresponding
kinetic energy of graphene for 2D wave vector {\bf
  k} is given by
$\epsilon_{{\bf k}s} = s \gamma |{\bf k}|$,
where $s=\pm 1$ indicate the conduction
(+1) and valence ($-1$) bands, respectively, and $\gamma$ is a
band parameter (and Fermi velocity of graphene is given by
$v_F=\gamma/\hbar$). The corresponding density of states (DOS) is
given by $ D(\epsilon) = g_s g_v |\epsilon|/(2\pi\gamma^2)$, where
$g_s=2$, $g_v=2$ are the spin and valley degeneracies,
respectively. The Fermi momentum ($k_F$) and the Fermi energy
($E_F$) of 2D graphene are given by $k_F = (4\pi n/g_s g_v)^{1/2}$
and $E_F = \gamma k_F$ where $n$ is the 2D carrier (electron or
hole) density.

Many electronic properties of a system are strongly
influenced by the presence of electron-electron interaction.
In this paper, we investigate theoretically the
electron-electron interaction induced exchange-correlation effects in
two dimensional graphene layer. We calculate, within the leading-order
dynamical-screening approximation, the electron self-energy for
extrinsic  graphene at zero temperature.
The self energy is the central quantity that determines the other
Fermi liquid parameters. We obtain the single particle
spectral function and  the
inelastic quasiparticle lifetimes and associated mean free paths from
the calculated electron self-energy.

The self-energy is given by in the screened interaction ($G_0W$
approximation) \cite{mahan}
\begin{eqnarray}
\Sigma_s({\bf k},i\omega_n) =
& - &\frac{1}{\beta}\sum_{s'}\sum_{{\bf q},i\nu_n}
G_{0,s'}({\bf k}+{\bf q},i\omega_n+i\nu_n)  \nonumber \\
&\times & \frac{V_{c}(q)}{\varepsilon({\bf q},i\nu_n)}
F_{ss'}({\bf k},{\bf k}+{\bf q})
\label{sigma}
\end{eqnarray}
where $\beta = 1/k_B T$, $s,s'=\pm 1$ denote the band indices,
$G_{0,s}({\bf k},i\omega_n)=1/(i\omega_n-\xi_{{\bf k}s})$ is the
unperturbed Green's function ($\xi_{{\bf k}s} =\epsilon_{{\bf k}s}
-\mu$ where $\mu$ is the chemical potential), $V_{c}(q)=2\pi
e^2/\kappa q$ is the bare Coulomb potential (with background
dielectric constant $\kappa$), $\varepsilon(q,\omega)$ is the
dynamical screening function (dielectric function) given by
$\varepsilon(q,\omega) = 1-v_c(q) \Pi(q,\omega)$, where
$\Pi(q,\omega)$, the 2D polarizability, is given by the bare
bubble diagram \cite{Hwang}. In Eq. (\ref{sigma}) $F_{ss'}({\bf
k},{\bf k}+{\bf q})$ is the overlap of states, given by
$F_{ss'}({\bf k},{\bf k}') = (1 + ss' \cos\theta_{{\bf
kk}'})/{2}$, where $\theta$ is the angle between ${\bf k}$ and
${\bf k}'$.

After the standard procedure of analytical continuation, the
self-energy can be separated into the exchange and correlation
parts $ \Sigma_{s}({\bf k},\omega) = \Sigma_{s}^{\rm ex}({\bf k})
+ \Sigma_{s}^{\rm cor}({\bf
  k},\omega). $
The exchange part is given by
\begin{equation}
\Sigma^{\rm ex}_{s}({\bf k}) = -\sum_{s'{\bf q}}
\ n_F(\xi_{{\bf k}+{\bf q},s'})\, V_c({\bf q})\, F_{ss'}({\bf
k}, {\bf k}+{\bf q}),
\label{eq:1}
\end{equation}
where $n_F(\xi_{{\bf k}s}) = \theta(\xi_{{\bf k}s})$ is the Fermi
function at $T=0$. The correlation part, $\Sigma^{\rm
cor}_{s}({\bf
  k},\omega)$, is defined to be
the part of $\Sigma_s({\bf k},\omega)$ not included in $\Sigma^{\rm ex}_{s}$.
Since the exchange part of self-energy $\Sigma^{\rm ex}_s({\bf
  k},\omega)$ is studied
in ref. \cite{exch} we provide in this paper the results of correlation part.
In the GW approximation, the $\Sigma^{\rm cor}_{s}({\bf k},\omega)$
can be written
in the line and pole decomposition $\Sigma^{\rm cor}_s = \Sigma^{\rm line}_s +
\Sigma^{\rm pole}_s$, where
\begin{eqnarray}
\Sigma^{\rm line}_s({\bf k},\omega) =
 - \sum_{s'{\bf q}}\int^{\infty}_{-\infty}
\frac{d\omega'}{2\pi} & &
\frac{V_c({\bf q})F_{ss'}({\bf k,k+q})}{\xi_{{\bf
      k+q},s'}-\omega-i\omega'} \nonumber \\
&\times& \left [
  \frac{1}{\epsilon(q,i\omega')} -1 \right]
\end{eqnarray}
\begin{eqnarray}
\Sigma^{\rm pole}_s({\bf k},\omega) &=& \sum_{s'{\bf q}}\left [
  \theta(\omega - \xi_{{\bf k+q},s'})-\theta(-\xi_{{\bf k+q},s'}) \right ]
\nonumber \\
&\times & V_c({\bf q})F_{ss'}({\bf k,k+q})
\left [ \frac{1}{\epsilon(q,\xi_{{\bf k+q},s'})}-1 \right ].
\end{eqnarray}
The $\Sigma_{\rm {line}}$ is completely real because
$\epsilon(q,i\omega)$ is real. Thus, Im[$\Sigma_{{\rm pole}}$] gives
the total contribution to the imaginary part of the self-energy.
In this calculation we use the cut-off at the wave vector $k_c$
with respect to the Dirac point; {\it i.e.}, the
integral is cut off at $k = k_c$.
From the calculated self energy $\Sigma(k,\omega)$, we can obtain the
single-particle spectral function given by

\begin{figure}[ht]
\begin{center}\leavevmode
\includegraphics[width=0.7\linewidth]{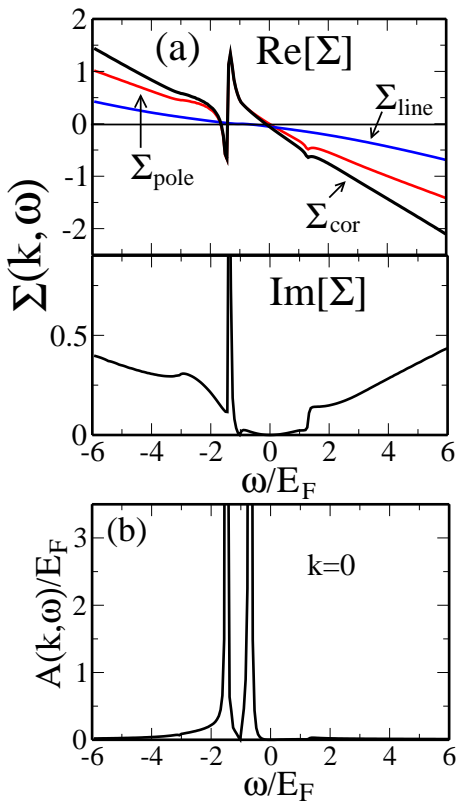}
\caption{(a) Real (top panel) and imaginary (bottom panel) parts of
  the self-energy $\Sigma(k,\omega)$ and (b) spectral function
  $A(k,\omega)$ as a function of the energy $\omega$ for $k=0$
  (the
  Dirac point).}
\label{fig1}
\end{center}
\end{figure}

\begin{equation}
A(k,\omega) = \frac{2{\rm Im}[\Sigma(k,\omega)]}{ \left \{
    \omega+\mu-\xi_k-{\rm Re}[\Sigma(k,\omega)] \right \}^2 + \left \{ {\rm
        Im}[\Sigma(k,\omega)] \right \}^2}.
\end{equation}
The spectral function $A(k,\omega)$ satisfies the sum rule
\begin{equation}
\int_{-\infty}^{\infty}\frac{d\omega}{2\pi}A(k,\omega) =1,
\end{equation}
which is generally satisfied to within less than a percent in our
numerical calculations.

\section{Results}

In Figs. 1 and 2 we show the real and the imaginary parts of the
self-energy and spectral function for $k=0$ (the Dirac point) and
$k=k_F$ (Fermi surface). Throughout this paper we use the
following parameters: dielectric constant $\kappa =2.5$, Fermi
velocity $v_F = \gamma/\hbar =10^8$ cm$/s$, and cut-off momentum
$k_c = 1/a$, where $a=2.46$ \AA \; is the lattice constant of
graphene.

\begin{figure}[ht]
\begin{center}\leavevmode
\includegraphics[width=0.7\linewidth]{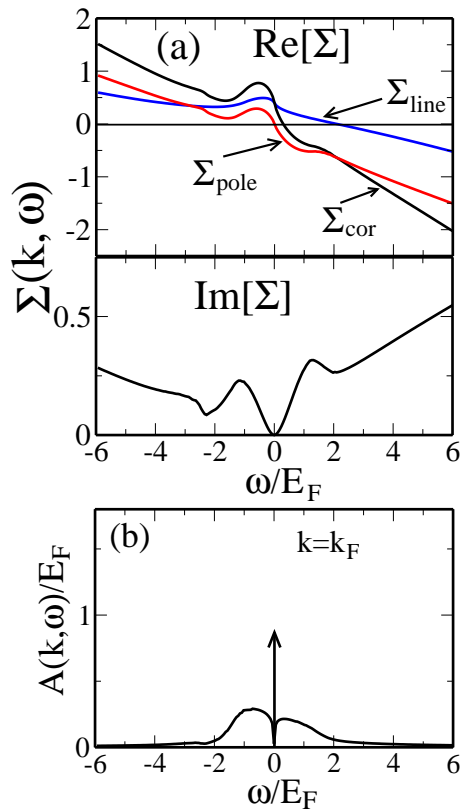}
\caption{(a) Real (top panel) and imaginary (bottom panel) parts of
  the self-energy $\Sigma(k,\omega)$ and (b) spectral function
  $A(k,\omega)$ as a function of the energy $\omega$ for $k=k_F$ (at the
  Fermi surface). The vertical arrow in (b) denote $\delta$-function
  in the spectral function of weight ($2\pi \times 0.9$) at $\omega=0$.}
\label{fig2}
\end{center}
\end{figure}

For $k=0$  (Fig. 1) we find a strong peak in Im[$\Sigma(k,\omega)$]
(associated with
a finite step in Re[$\Sigma(k,\omega)$]) due to plasmon emission.
Since Im[$\Sigma$] is finite for all energies except $\omega=0$ we expect
a finite damping in the spectral function. In spectral function we
find two peaks.
The first peak (from $\omega=0$) in $A(k,\omega)$ corresponds
to the usual
quasiparticle (i.e. a bare particle surrounded by a cloud of virtual
plasmons and particle-hole excitations), and the second peak corresponds
to a plasmaron, which is interpreted as a hole coupled to a cloud
of real plasmons \cite{jalabert}.

In Fig. 2 we show the self-energy and spectral function for the
Fermi wave vector ($k=k_F$). In particular, the behavior of the
self-energy and spectral function at $k=k_F$ determines the
low-energy properties of the system. For $k=k_F$  there is only
one peak in $A(k,\omega)$ at $\omega=0$, which is a
$\delta$-function peak because Im$\Sigma(k_F,\omega) \propto
\omega^2 \ln |\omega|$ as $\omega \rightarrow 0$ \cite{dassarma}.
Thus, the doped (or gated) graphene is a Fermi liquid because  it
possesses a Fermi surface (or a discontinuity in momentum
distribution function) whose presence is indicated by a $\delta$
function in $A(k_F,\omega)$ at $\omega=0$ (or, equivalently,
non-zero many-body renormalization factor).

In Fig. 3 we show  a  quasiparticle
lifetime due to electron--electron interactions, which is given by
$1/\tau (k) = 2 {\rm Im}[\Sigma(k,\xi_k)]$. Since the velocity of
the quasiparticles close the the Dirac points is approximately a
constant, the inelastic mean free path $\ell$ is obtained by
$\ell(\xi) = v_F \tau(\xi)$.  In inset of Fig.~3, we provide the
corresponding $\ell$, which shows that at $n=10^{13}\,{\rm cm}^{-2}$ a
hot electron injected with an energy of 1\,eV above
$E_F$ has an $\ell$ due to electron--electron interactions that is on
the order of 20\,nm. This will have implications for designing any hot
electron transistor type graphene devices.
Note that the calculate scattering time
is a smooth function because both plasmon emission and
interband processes are absent.
In parabolic band semiconductors \cite{jalabert} plasmon emission and
interband collision thresholds cause discontinuities in the scattering
time.

\begin{figure}[ht]
\begin{center}\leavevmode
\includegraphics[width=0.7\linewidth]{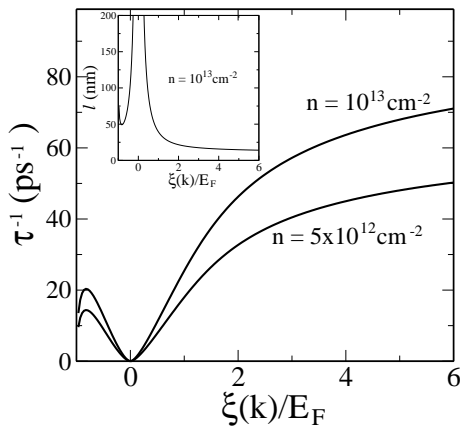}
\caption{
Inelastic quasiparticle lifetime,
$1/\tau=2{\rm Im}\Sigma(k,\xi_k)$, in graphene due to dynamically screened
electron--electron interactions,
as a function of energy at $T=0$ for different densities. Inset shows
the corresponding quasiparticle mean
free path for $n=10^{13}$ cm$^{-2}$.
}
\label{mfp}
\end{center}
\end{figure}

\section{Summary}

In summary, we calculate the electron-electron interaction induced
quasiparticle self-energy in graphene layer within the
leading-order dynamical-screening approximation. Our calculated
spectral function indicates that the extrinsic graphene (i.e. $E_F
\neq 0$) is a Fermi liquid. We also calculate hot carrier
inelastic scattering due to electron-electron interactions. The
linear dispersion and chiral property of graphene give lifetime
energy dependences that are qualitatively different from those of
parabolic-band semiconductors. We can apply our calculated
self-energy to obtain other quasi-particle properties of graphene.

This work is supported by US-ONR and LPS-NSA.

\end{document}